\documentclass[12pt]{article}

\font\grande=cmr9.5 scaled \magstep4
\font\medio=cmr9.5 scaled \magstep2
\outer\def\beginsection#1\par{\medbreak\bigskip
      \message{#1}\leftline{\bf#1}\nobreak\medskip
\vskip-\parskip
      \noindent}
\usepackage{graphicx} 
\begin{document}
\bibliographystyle {unsrt}

\titlepage

\begin{flushright}
CERN-PH-TH/2013-152
\end{flushright}

\vspace{10mm}
\begin{center}
{\grande Anomalous Magnetohydrodynamics}\\
\vspace{1.5cm}
 Massimo Giovannini
 \footnote{Electronic address: massimo.giovannini@cern.ch}\\
\vspace{1cm}
{{\sl Department of Physics, 
Theory Division, CERN, 1211 Geneva 23, Switzerland }}\\
\vspace{0.5cm}
{{\sl INFN, Section of Milan-Bicocca, 20126 Milan, Italy}}
\vspace*{0.5cm}
\end{center}

\vskip 0.5cm
\centerline{\medio  Abstract}
Anomalous symmetries induce currents which can be parallel rather than orthogonal 
to the hypermagnetic field. Building on the analogy with charged liquids at high magnetic Reynolds numbers, 
the persistence of anomalous currents is scrutinized for parametrically large conductivities when the plasma approximation is accurate. 
Different examples in globally neutral systems suggest that the magnetic configurations minimizing the 
energy density with the constraint that the helicity be conserved coincide, in the perfectly conducting limit, with the ones obtainable in 
ideal magnetohydrodynamics where the anomalous currents are neglected. It is argued that this is the rationale 
for the ability of extending to anomalous magnetohydrodynamics the hydromagnetic solutions characterized by finite gyrotropy.
The generally covariant aspects of the problem are addressed with particular attention to conformally flat geometries which are potentially relevant for the description of the electroweak plasma prior to the phase transition. 

\vskip 0.5cm

\noindent

\vspace{5mm}

\vfill
\newpage
\renewcommand{\theequation}{1.\arabic{equation}}
\setcounter{equation}{0}
\section{Introduction}
\label{sec1}
There are physical situations were electric currents are directed 
along the magnetic field itself.  For instance, in the analysis of ordinary hydromagnetic 
nonlinearities it is customary to study the evolution of the magnetic field averaged over the turbulent 
flow, be it compressible (as in acoustic turbulence) 
or incompressible \cite{kaza,kraichnan}. In the latter case the effective current density is proportional 
to the magnetic field as established in different situations 
and extensively reviewed in a number of textbooks \cite{moffat,parker,zeldovich}.
For this to happen a necessary condition is the parity breaking associated 
with the turbulent velocity field which has to be globally non-mirror-symmetric for sufficiently large 
kinetic Reynolds numbers. In this situation the averaged scalar product of the bulk velocity of the plasma 
with the bulk vorticity (sometimes called kinetic gyrotropy \cite{kaza,zeldovich}) does not vanish (i.e. 
$\langle \vec{v} \cdot (\vec{\nabla} \times \vec{v}) \rangle \neq 0$) and the kinetic energy of the plasma can be in principle transferred  to 
the magnetic field.

Provided pseudoscalar species exist in the plasma, the effective Ohmic currents can be oriented along the magnetic field 
even if turbulent flows are absent. For instance, in the case of axions \cite{kim,raffelt,PQ} the standard model 
is supplemented by a (global) $U_{PQ}(1)$. This symmetry is broken at the Peccei-Quinn scale $F_{a}$ and leads to a  dynamical pseudo Goldstone boson (the axion) presumably acquiring a small mass because of soft instanton effects 
at the QCD phase transition. If an axionic density is present in the early 
Universe, bounds can be obtained for the Peccei-Quinn symmetry breaking scale.
These bounds together with other constraints leave a window opportunity 
$ F_{a} \simeq {\mathcal O} (10^{10})\, {\rm GeV}$ with many uncertainties concerning the axion mass \cite{raffelt}.
Pseudoscalar species can also arise in the low energy limit of superstring models but 
in spite of its specific physical origin, the pseudoscalar field (be it $\psi$) 
can couple to the Abelian gauge field strength $Y_{\mu\nu}$ as $(\psi/M) Y_{\mu\nu}\,\tilde{Y}^{\mu\nu}$
where $\tilde{Y}^{\mu\nu}$ is the dual field strength and $M$ is related, in the axion case, to the Peccei-Quinn scale. 

In the symmetric phase of the electroweak theory the hypercharge current can flow along the 
hypermagnetic field. Both the current and the magnetic field 
are usual vector fields and the proportionality factor is related to the chemical potential of the anomalous charges.
This effect arises in gauge theories at finite density where it can happen that cold fermionic matter with 
non-zero anomalous Abelian charges is unstable against the creation of Abelian gauge field \cite{rub,red}.
The existence of currents parallel to the Abelian gauge field strength has been also analyzed in the electroweak 
plasma \cite{mg} with the aim of understanding how hypercharge fields may be converted into fermions in a  hot environment.
A magnetic field intensity parallel (or antiparallel) to the current density is also thought to be one of the 
potential consequences of the existence of the quark-gluon plasma and it has been more recently 
studied in the context of heavy ions collisions \cite{kh} as well as in holographic approaches \cite{hol}.
Chiral anomalies alter the evolution of the corresponding current but also the evolution of the gauge fields and of 
the corresponding Ohmic currents. This is point common to all the themes mentioned in this paragraph.

In the present investigation the persistence of anomalous currents will be scrutinized in globally neutral and conducting plasmas  at high temperatures such as the ones occurring prior to matter radiation equality or in the symmetric phase of the electroweak theory. The terminology anomalous magnetohydrodynamics (AMHD in what follows) refers to the evolution of hydromagnetic nonlinearities in the presence of anomalous symmetries both in  cold and hot environments. In section \ref{sec2}  the 
case of ordinary hydromagnetic nonlinearities shall be reviewed and some basic terminology will be introduced.
Section \ref{sec3} is devoted to the birefringence induced by the axial couplings with the aim of 
deriving the evolution of the slow modes of the globally neutral and conducting plasma in the simplified 
situation where the the total energy-momentum tensor of the system is covariantly conserved. 
In section \ref{sec4} we shall move to the case where there are two currents one anomalous and the 
other conducting always under the assumption of the global neutrality of the plasma. It will be shown that the second law 
of thermodynamics constrains the conduction current which must contain both magnetic and vortical components.
In section \ref{sec5} the 
ideal and the resistive limits of AMHD will be studied in the case of a conformally flat background geometry. Section \ref{sec6} contains the concluding remarks. 
In the appendix various results shall be swiftly derived with the aim of easing the derivations
presented in the bulk of the paper. Unlike previous analyses (bounded 
to a special relativistic treatment)  we shall privilege here the generally covariant approach which is more suitable 
for the applications to curved space-times and, more specifically, to conformally flat background geometries.

\renewcommand{\theequation}{2.\arabic{equation}}
\setcounter{equation}{0}
\section{Hydromagnetic nonlinearities}
\label{sec2}
Magnetohydrodynamics (MHD in what follows) can be investigated within two different but in some sense complementary approaches. The 
ideal (or perfectly conducting) limit where  the conductivity goes to infinity (i.e. the $\sigma_{\mathrm{c}} \to \infty$ limit) and  the real (or resistive) limit where the conductivity is finite (see, for instance, Refs. \cite{freidberg,biskamp,principles}). The ordinary magnetic diffusivity equation in ideal MHD can be simply written as:
\begin{equation}
\frac{\partial \vec{B}}{\partial \tau} = \vec{\nabla} \times (\vec{v} \times \vec{B}) + {\mathcal O}(\sigma_{\mathrm{c}}^{-1}), 
\label{D1}
\end{equation}
where $\vec{v}$ denotes the bulk velocity of the plasma and $\vec{B}$ is the magnetic field 
intensity.  Batchelor  \cite{batchelor} pioneered the general picture of the interaction between the magnetic field and a conducting liquid by exploiting the analogy with a bulk velocity vortex in an incompressible liquid. 
Assuming, as often done in statistical fluid mechanics, that the bulk velocity of the charged fluid is
stationary and isotropic, the correlation function of the velocity field can be written as \cite{kaza,kraichnan}:
\begin{equation}
\langle{v}_{i}(\vec{k},\tau)\, v_{j}(\vec{p},\tau')\rangle = \biggl[{\mathcal A}_{1}(k) P_{ij}(\hat{k}) + {\mathcal A}_{2}(k) \epsilon_{ijk} \hat{k}^{k}\biggr]\, \delta^{(3)}(\vec{k}+ \vec{p})\,f(\tau,\tau'),
\label{D2}
\end{equation}
where $P_{ij}(\hat{k}) = (\delta_{ij} - \hat{k}_{i} \hat{k}_{j}) $ and $\hat{k}_{i} = k_{i}/k$.
In the Markovian approximation $f(\tau,\tau')$ is proportional to $\delta(\tau-\tau')$ and the power spectra can have different forms which are not immediately 
relevant for the present considerations. Using Eqs. (\ref{D2}) and Eq. (\ref{D1}) the effective evolution equation for the magnetic field 
averaged over the bulk velocity field is
\begin{equation}
\frac{\partial \vec{H}}{\partial\tau} = \alpha \vec{\nabla} \times \vec{H}, \qquad \alpha = - \frac{\tau_{\mathrm{c}}}{3} \langle \vec{v} \cdot (\vec{\nabla}\times \vec{v}) \rangle.
\label{D3}
\end{equation}
Since the ideal hydromagnetic limit is a slow description valid for large distances,
 the displacement current can be neglected so that $\vec{\nabla} \times\vec{H}$ is proportional to the current density $\vec{j}$.
 But then Eq. (\ref{D3}) implies that there is an effective Ohmic current proportional to $\vec{H}$ 
\cite{zeldovich}.  In the Zeldovich interpretation \cite{zeldovich,zeldovich2}, Eq. (\ref{D3}) suggests that an ensemble of screw-like vortices with
zero mean helicity is able to generate loops in the magnetic flux. Equations (\ref{D2}) and (\ref{D3}) have been analyzed for a number of astrophysical applications and describe the physical situation where kinetic energy is transferred to magnetic energy.

The plasma description following from MHD can be also phrased in terms of the conservation of two interesting quantities, i.e. 
the magnetic flux and the magnetic helicity  \cite{freidberg,biskamp,principles}
\begin{eqnarray}
&& \frac{d}{d\tau} \int_{\Sigma} \vec{B} \cdot d\vec{\Sigma}=- \nu_{\mathrm{mag}} \int_{\Sigma} \vec{\nabla} \times(\vec{\nabla}
\times\vec{B})\cdot d\vec{\Sigma},
\label{FLCONS}\\
&&\frac{d}{d \tau}\int_{V} d^3 x \, \vec{A}~\cdot \vec{B} = - 2 \nu_{\mathrm{mag}} \int_{V} d^3 x
{}~\vec{B}\cdot(\vec{\nabla} \times\vec{B}),
\label{HELCONS}
\end{eqnarray}
where $V$ is a fiducial volume comoving with the conducting fluid and $\Sigma$ is the corresponding boundary surface; we defined $\nu_{\mathrm{mag}} = 1/(4\pi \sigma_{\mathrm{c}})$. Up to a gauge coupling constant the magnetic helicity coincides with the Chern-Simons number. The 
quantity $\vec{B}\cdot(\vec{\nabla} \times\vec{B})$ is sometimes called magnetic gyrotropy in full analogy with the kinetic gyrotropy 
already mentioned in the introduction. 

In a conducting plasma the kinetic and magnetic Reynolds numbers are defined as $R_{\mathrm{kin}} = v_{\mathrm{rms}}\, L_{v}/\nu_{\mathrm{kin}}$ and  $R_{\mathrm{mag}} = v_{\mathrm{rms}}\, L_{B}/\nu_{\mathrm{mag}}$ where $v_{\mathrm{rms}}$ estimates the bulk velocity of the plasma while $\nu_{\mathrm{kin}}$ denotes the coefficient of thermal diffusivity; $L_{v}$ and $L_{B}$ are, respectively, the correlation scales of the velocity field and of the magnetic field.   In the 
ideal hydromagnetic limit (i.e. $\sigma_{\mathrm{c}} \to \infty$, $\nu_{\mathrm{mag}} \to 0$ and $R_{\mathrm{mag}}\to \infty$)
the flux is exactly conserved and the number of links and twists in the magnetic flux lines is also preserved by the time evolution.  If $R_{\mathrm{kin}} \gg 1$ and  $R_{\mathrm{mag}} \leq {\mathcal O}(1)$  the system is still turbulent; however, since the total time derivative of the magnetic flux and of the magnetic helicity are both ${\mathcal O}(\nu_{\mathrm{mag}})$ the terms at the right hand side of Eqs. (\ref{FLCONS})--(\ref{HELCONS}) cannot be neglected.  Finally,  
if $R_{\mathrm{mag}} \gg 1$  and $R_{\mathrm{kin}} \ll 1$ the fluid is not kinetically turbulent but the magnetic flux 
is conserved. This occurs, incidentally, after matter radiation equality but before decoupling \cite{mg2}.
The considerations developed here are bound to the analysis of a number of toy models but they are potentially relevant in more realistic 
situations as long as the plasma can be considered globally neutral and perfectly conducting. First-order phase transitions, if they occur in the early Universe,  can provide a source of kinetic turbulence and, hopefully, the possibility of inverse cascades which could lead to an enhancement of the correlation scale of a putative large-scale magnetic field \cite{PT}. The extension of the viewpoint conveyed in the present analysis to kinetically turbulent environment is not implausible but shall not be attempted here. For the present ends what matters are the physical analogies of the 
forthcoming discussions with the physics of charged liquids at high magnetic reynolds numbers.

\renewcommand{\theequation}{3.\arabic{equation}}
\setcounter{equation}{0}
\section{Dynamical pseudoscalar fields}
\label{sec3}
Turbulence at high Reynolds numbers is sufficient for the existence of Ohmic currents flowing, in average, along the magnetic field direction. 
Such a requirement is, however, not necessary since similar phenomena can arise thanks to pseudoscalar species. 
Denoting with 
$S_{\psi}$ the pseudoscalar contribution and  with $S_{Y}$ the gauge part, the corresponding actions can be written as\footnote{The conventions will be the following. Greek indices run over the four-dimensional space-time. Latin (lowercase) indices run over three-dimensional spatial geometry. The signature 
of the metric is mostly minus, i.e. $(+,\,-,\, -,\,-)$.}:
\begin{eqnarray} 
S_{\psi} + S_{Y} &=& \int d^{4} x \sqrt{-g} \biggl[ \frac{1}{2}g^{\alpha\beta} 
\partial_{\alpha}\psi \partial_{\beta} \psi - W(\psi)  - j^{\alpha} Y_{\alpha} \biggr] 
\nonumber\\
&-& \frac{1}{16\pi} \int d^{4} x \sqrt{- g} \biggl[ Y_{\alpha\beta} Y^{\alpha\beta} + 
\frac{\psi}{M} Y_{\alpha\beta} \tilde{Y}^{\alpha\beta} \biggr],
\label{action}
\end{eqnarray}
where $j^{\alpha} = j^{\alpha}_{(+)}  - j^{\alpha}_{(-)}$ and $j_{\pm}^{\alpha} = \tilde{n}_{\pm} \, u_{\pm}^{\alpha}$; the velocities $u_{(\pm)}^{\alpha}$ satisfy $g_{\alpha\beta} u_{(\pm)}^{\alpha}\,u_{(\pm)}^{\beta} =1$ and $Y_{\alpha\beta}$ is the gauge field strength. The equations for $\psi$  and $Y^{\mu\nu}$ are obtained by minimizing the variation of  the action (\ref{action}) and they are:
\begin{eqnarray}
&& g^{\alpha\beta} \nabla_{\alpha} \nabla_{\beta} \psi + \frac{\partial W}{\partial \psi} = - \frac{1}{16\pi M} 
Y_{\alpha\beta} \tilde{Y}^{\alpha\beta},
\label{eqpsi}\\
&&  \nabla_{\alpha} Y^{\alpha\beta} = 4 \pi j^{\beta} - \frac{\partial_{\alpha} \psi}{M} \tilde{Y}^{\alpha\beta},
\label{eqY}
\end{eqnarray}
where $\nabla_{\alpha}$ denotes the covariant derivative.
The exchange of energy and momentum between the charged species is responsible for the existence of a finite conductivity.
The presence of an energy-momentum transfer $\Gamma$ implies:
\begin{eqnarray}
&&\nabla_{\mu} T^{\mu\nu}_{(+)} + \Gamma g^{\alpha\nu} (p_{-} + \rho_{-}) u_{\alpha} = Y^{\nu\alpha} j^{(+)}_{\alpha}, 
\label{cov1}\\
&& \nabla_{\mu} T^{\mu\nu}_{(-)} - \Gamma g^{\alpha\nu} (p_{-} + \rho_{-}) u_{\alpha} = -Y^{\nu\alpha} j^{(-)}_{\alpha},
\label{cov2}
\end{eqnarray}
where $u_{\alpha}$ denotes the total velocity field and $T^{\mu\nu}_{(\pm)}$ is:
\begin{equation}
T^{\mu\nu}_{(\pm)} = (p_{\pm} + \rho_{\pm}) u^{\mu}_{(\pm)} u^{\nu}_{(\pm)}  - p_{\pm} g^{\mu\nu}. 
\label{cov3}
\end{equation}
The relation between $u^{\mu}$ and $u^{\nu}_{(\pm)}$ is: 
\begin{equation}
u^{\mu} u^{\nu} = \biggl(1 + \gamma_{+}\biggr) \Omega_{+} u_{(+)}^{\mu} u_{(+)}^{\nu} + \biggl(1 + \gamma_{-}\biggr) \Omega_{-} u_{(-)}^{\mu} u_{(-)}^{\nu}, 
\label{cov4}
\end{equation}
where $\gamma_{\pm} = p_{\pm}/\rho_{\pm}$ and $\Omega_{\pm} = \rho_{\pm}/(\rho_{+} + \rho_{-})$; note also that $g_{\alpha\beta} u^{\alpha}u^{\beta} =1$.
Equations (\ref{cov1}) and (\ref{cov2}) can be summed and subtracted. From the sum we get the equation for the 
total energy-momentum tensor of the charges, i.e.
\begin{equation}
\nabla_{\mu} T^{\mu\nu}(\rho, p) = Y^{\nu\alpha} j_{\alpha}, \qquad j_{\alpha} = j_{\alpha}^{(+)} - j_{\alpha}^{(-)}.
\label{sum}
\end{equation}
From the difference of Eqs. (\ref{cov1}) and (\ref{cov2}) (multiplied by the corresponding charge concentrations) an evolution 
equation for the total current can be obtained. In the limit where the rate of interaction dominates against the rate 
of variation of the geometry this combination leads to a relation between the current and the gauge field 
strength i.e. the Ohm law which will be introduced in a moment. 

For the subsequent applications it is useful to rephrase the evolution of the system in terms of the evolution of the energy-momentum tensors
for $T_{\mu}^{\nu}(\psi)$ and $T_{\mu}^{\nu}(Y)$:
\begin{eqnarray}
\nabla_{\mu} T^{\mu}_{\nu}(\psi) = - \frac{\partial_{\nu} \psi}{16\pi M} Y_{\alpha\beta} \tilde{Y}^{\alpha\beta},
\label{cov5}\\
\nabla_{\mu} T^{\mu}_{\nu}(Y) = - Y_{\nu\alpha}\, j^{\alpha}  + \frac{\partial_{\nu} \psi}{16\pi M} Y_{\alpha\beta} \tilde{Y}^{\alpha\beta},
\label{cov6}
\end{eqnarray}
where 
\begin{eqnarray}
&& T_{\mu}^{\nu}(\psi) = \partial_{\mu}\psi \partial^{\nu}\psi - \delta_{\mu}^{\nu} \biggl[ \frac{1}{2} g^{\alpha\beta} \partial_{\alpha}\psi 
\partial_{\beta} \psi - W(\psi)\bigg],
\label{tens1}\\
&& T_{\mu}^{\nu}(Y) = \frac{1}{4\pi}\biggl[ - Y_{\mu\alpha} Y^{\nu\alpha} + \frac{\psi}{M}  Y_{\mu\alpha} \tilde{Y}^{\nu\alpha}
+ \frac{1}{4} \delta_{\mu}^{\nu} \biggl( Y_{\alpha\beta} Y^{\alpha\beta} + Y_{\alpha\beta} \tilde{Y}^{\alpha\beta}\biggr) \biggr].
\label{tens2}
\end{eqnarray}

Consider now a conformally flat geometry of the type 
$g_{\mu\nu} = a^2(x) \eta_{\mu\nu}$ where $\eta_{\mu\nu}$ is the Minkowski metric and the scale factor can be a function of a generic 
space-time point. The gauge field strengths 
can be written as $Y_{i0} = - a^2(x)\,e_{i} $ and $Y_{ij} = - a^2(x) b^{k} \epsilon_{ijk}$ and the
equations for the hyperelectric and hypermagnetic fields are given, in this case, by:
\begin{eqnarray}
&& \vec{\nabla} \cdot \vec{E} = 4 \pi ( n_{+} - n_{-} ) 
- \frac{1}{M} \vec{\nabla}\psi\cdot \vec{B},
\label{div1}\\
&&\vec{\nabla} \cdot \vec{B} =0,\qquad \vec{\nabla} \times \vec{E} = -\partial_{\tau} \vec{B},
\label{bianchi1}\\
&& \vec{\nabla} \times  \vec{B} = \partial_{\tau} \vec{E} + \frac{1}{M} 
\biggl[\partial_{\tau}\psi \vec{B} + \vec{\nabla}\psi \times \vec{E} \biggr] 
 + 4 \pi [ n_{+} \vec{v}_{+} - n_{-} \vec{v}_{-} ],
\label{current}
\end{eqnarray}
where $\vec{v}_{\pm} = a \vec{u}_{\pm}$; furthermore $\vec{E} = a^2 \vec{e}$ and $\vec{B} = a^2 \vec{b}$. The expressions of the 
total charge and current density appearing in Eqs. (\ref{div1}) and (\ref{current}) come from the definition 
of the comoving concentrations of charged species (i.e. $n_{\pm}(x) = a^3(x) \tilde{n}_{\pm}$) and from the  
comoving velocity field (i.e. $\vec{v}_{\pm} = a \vec{u}_{\pm}$).
The covariant conservations of the currents leads to the 
evolution equations of the comoving concentrations, i.e. $\partial_{\tau} n_{\pm} + \vec{\nabla}\cdot( n_{\pm} \vec{v}_{\pm}) =0$.

For the present ends the interesting situation contemplates a globally neutral plasma.
From Eq. (\ref{div1})  $\vec{\nabla}\cdot \vec{E} =0$ provided  $n_{+} = n_{-} = n_{0}$ and, at the same time, $\psi$ is spatially homogenous, i.e. $\vec{\nabla} \psi =0$. 
In the limit where the rate of interaction between the charged species is larger than the rate 
of variation of the geometry the total Ohmic current can be expressed in covariant language as \cite{lic}
$j^{\alpha} =\sigma_{\mathrm{c}}\, Y^{\nu\alpha} \,u_{\alpha}$ where $\sigma_{\mathrm{c}}$ is the conductivity of the system. From the expression 
of the Ohmic current $j^{\alpha} u_{\alpha} =0$ and this is why we can also write  $j^{\mu} h_{\mu}^{\nu} = \sigma_{\mathrm{c}} Y^{\nu\alpha} u_{\alpha}$. If we now project the latter expression along $u_{\nu}$ we obtain an identity. Conversely, if we project $j^{\mu} h_{\mu}^{\nu}$ along $h_{\nu}^{\beta}= (\delta_{\nu}^{\beta} - u_{\nu} u^{\beta})$ we shall obtain, again, $j^{\alpha} =\sigma_{\mathrm{c}}\, Y^{\nu\alpha} \,u_{\alpha}$ since, by definition,  $h_{\mu}^{\nu} \, h_{\nu}^{\beta} = h_{\mu}^{\beta}$.

The value of the conductivity depends on the specific properties of the plasma. In particular, defining with $m$ the mass 
of the lightest charge carrier we have that $\sigma_{\mathrm{c}}\simeq T/\alpha_{\mathrm{em}}$ for $T\gg m$ and $ \sigma_{\mathrm{c}} \simeq (T/\alpha_{\mathrm{em}}) \,(T/m)^{1/2}$ in the opposite limit. The total Ohmic current is then given by:
\begin{equation}
\vec{J} = \sigma \biggl(\vec{E} + \vec{v}\times \vec{B} + \frac{\vec{\nabla} p}{n_{0}}- \frac{\vec{J}\times \vec{B}}{n_{0}}\biggr),
\label{APP4}
\end{equation}
where $n_{0}$ denotes the rescaled charge concentration while $\sigma = \sigma_{\mathrm{c}} a$ and $\vec{J} = a^3 \vec{j}$. Dropping the third 
and fourth terms at the right hand side of Eq. (\ref{APP4}) (i.e. the thermoelectric and Hall terms which are of higher order in the spatial gradients)  the (hyper)electric field can be expressed in terms of the total current.  

In the resistive approximation, the hyperelectric field is not exactly orthogonal to the hypermagnetic one. The source 
of this mismatch depends on the specific dynamical situation and, in the present case, the induced hyperelectric field is:
 \begin{equation}
{\vec{E}}
 \simeq \frac{\vec{\nabla}\times \vec{B}}{4\pi \sigma} - \frac{\partial_{\tau} \psi \vec{B} - \vec{\nabla}\psi \times (\vec{v} \times \vec{B})}{4 \pi \sigma} - \vec{v} \times \vec{B} + {\mathcal O}(\sigma^{-2})
\label{induced}
\end{equation}
The first term at the right hand side of Eq. (\ref{induced})  is the analog of the  MHD
contribution. The second and third contributions contain both temporal and spatial derivatives of $\psi$ and 
describe the energy-momentum transfer from the pseudoscalar field to the 
 hypermagnetic field. Depending on the initial topology of the hypermagnetic flux lines this process can even produce 
 hypermagnetic knots (see \cite{mg} third and fourth papers). In Eq. (\ref{induced}) there are various other terms 
 proportional to the gradients of $\psi$ and carrying terms ${\mathcal O}(\sigma^{-2})$, ${\mathcal O}(\sigma^{-3})$ and so on and so forth.
Using Eq. (\ref{induced}) into Eq. (\ref{bianchi1}) we obtain the wanted form of the magnetic diffusivity equation:
\begin{eqnarray}
\frac{\partial \vec{B}}{\partial \tau} = \frac{\vec{\nabla}\times (\partial_{\tau} \psi\, \vec{B})}{4\pi M \sigma} - \frac{\vec{\nabla}\times[(\vec{\nabla}\psi) \times (\vec{v} \times\vec{B})]}{4\pi \sigma M}+ \vec{\nabla}\times (\vec{v}\times \vec{B}) + \frac{\nabla^2 \vec{B}}{4 \pi \sigma}.
\label{mdiff}
\end{eqnarray} 
Equation (\ref{mdiff}) generalizes Eq. (\ref{D2}) and the first term at the right 
hand side can be interpreted as a current density flowing along the magnetic field.
From Eq. (\ref{mdiff}) it can be immediately argued that whenever the conductivity is high (and the ideal limit can be 
enforced) the magnetic current is suppressed by the value of the conductivity. 

The total energy-momentum tensor $T_{\mathrm{tot}}^{\mu\nu}$  
is covariantly conserved, i.e. $\nabla_{\mu} T^{\mu\nu}_{\mathrm{tot}} =0$ as it can be 
easily argued by combining Eqs. (\ref{cov1})--(\ref{cov2}) and (\ref{cov5})--(\ref{cov6}). 
Also the total entropy of the system is covariantly conserved. However, recalling the first principle 
of thermodynamics and the fundamental thermodynamic identity, it can be easily shown that 
the entropy of the global fluid of charged species obeying Eq. (\ref{sum}) is not conserved and the corresponding 
evolution equation of the entropy four-vector is:
\begin{equation}
\nabla_{\mu} \varsigma_{\mathrm{f}}^{\mu} = \frac{\sigma}{T} Y_{\alpha\beta} Y^{\nu\alpha} \,u_{\nu}\,u^{\beta}, \qquad \varsigma_{\mathrm{f}} = \frac{\rho + p}{\overline{T}},
\label{prec}
\end{equation}
where $\varsigma_{\mathrm{f}}^{\mu} = \varsigma_{\mathrm{f}}\, u^{\mu}$ and the term appearing at the right hand side 
of the conservation equation is noting but the relativistic generalization of the heating due to Joule effect; $\overline{T} = a T$ denotes the 
comoving temperature. When the conductivity vanishes gauge fields can be amplified thanks to the coupling to the pseudoscalar field and this phenomenon 
has been studied in various space-times and, in particular, during a quasi-de Sitter 
stage of expansion \cite{carroll,bamba,campanelli} (see also third and fourth papers in \cite{mg}).
If we start with a field configuration carrying zero magnetic helicity the pseudoscalar 
coupling discussed here can produce configurations characterized by non-vanishing 
magnetic helicity which have been dubbed hypermagnetic knots. 

\renewcommand{\theequation}{4.\arabic{equation}}
\setcounter{equation}{0}
\section{Anomalous symmetries at finite density}
\label{sec4}
The derivation of the previous section assumed the global neutrality of the plasma and the covariant conservation 
of the total energy-momentum tensor. The latter approach shall now be reversed by dealing directly with the currents rather than with a 
specific form of the action. In this respect the simplified model discussed in the present section is instructive insofar as it contemplates the simultaneous presence of two currents one anomalous and the other non-anomalous (to be identified, in the language of the previous section, with the hyperelectric current). The logic is, in short, the following. The anomalous current $j^{\mu}_{R}$ is not covariantly conserved because of the anomaly contribution:
\begin{equation}
\nabla_{\mu} j^{\mu}_{R} = {\mathcal A}_{R} Y_{\mu\nu} \tilde{Y}^{\mu\nu}.
\label{an1}
\end{equation}
The equations of the energy-momentum tensor of the fluid and the covariant conservation of the non-anomalous four-current are instead:
\begin{equation}
\nabla_{\mu} T^{\mu\nu}  = Y_{\nu\alpha}\, j^{\alpha}, \qquad \nabla_{\mu} j^{\mu} =0.
\label{an2}
\end{equation}
The equation for  $T^{\mu\nu}$ appearing in Eq. (\ref{an2}) can be split
in terms of the two projections along $u^{\nu}$ and along $h^{\nu}_{\alpha} = \delta^{\nu}_{\alpha} - u^{\nu} u_{\alpha}$ (see Eqs. (\ref{cons1})--(\ref{cons2}) of appendix \ref{APPA}). 
The system of Eqs. (\ref{an1})--(\ref{an2}) approximately describes different physical situations ranging from the anomalous plasma 
in the symmetric phase of the electroweak theory \cite{mg}, to the models of chiral liquid \cite{kh} which are 
proposed as a simplified framework for the discussion of the quark gluon plasma. 
Indeed above the critical temperature of the corresponding phase transition the electroweak symmetry is
restored, and the non-screened gauge field strength $Y_{\mu\nu}$ corresponds to the
U(1)$_Y$ hypercharge group.  The system of Eqs. (\ref{an1}) and (\ref{an2}) extends the hydrodynamic approach described in Ref. \cite{SS} 
to the extent that $j^{\mu}$ does not coincide with $j^{\mu}_{R}$ and the ambient plasma is globally neutral. As already mentioned, unlike previous analyses bounded 
to a special relativistic treatment,  the generally covariant discussion is more suitable 
for the class of problems addressed here.

\subsection{Useful thermodynamic relations} 
Denoting with $\mu_{R}$ the chemical potential associated with the anomalous species, the first principle of thermodynamics implies:
\begin{equation}
d E = T d S - p d V + \mu_{R} d N_{R}.
\label{th1}
\end{equation}
Dividing the fundamental thermodynamics identity (i.e. $E = T S - p V + \mu_{R} N_{R}$) by a fiducial 
volume we obtain the well known relation $\rho + p = T \varsigma + \mu_{R} \tilde{n}_{R}$. Differentiating
the fundamental thermodynamic identity and subtracting the obtained result from Eq. (\ref{th1})
a known relation between the ordinary derivatives of the temperature, of the chemical potential and of the pressure 
can be obtained and it is, in the present case, 
\begin{equation}
\varsigma \partial_{\alpha} T + \tilde{n}_{R} \partial_{\alpha} \mu_{R} = \partial_{\alpha} p.
\label{th1a}
\end{equation} 
The anomalous current of Ref. \cite{mg} was associated with the slowest perturbative processes related to the $U(1)_{Y}$ anomaly, 
namely the processes flipping the chirality of the right electron which are in thermal equilibrium until sufficiently late because 
of the smallness of their Yukawa coupling. 
The origin of the anomalous current is not essential for the present ends but what matters is the physical and mathematical 
distinction between anomalous and conduction (possibly Ohmic) currents. According to this approach, the general expression of the anomalous current must contemplate  an inviscid contribution supplemented by a viscous term, i.e. 
$j_{R}^{\mu} = \tilde{n}_{R} u^{\mu} + \nu_{R}^{\mu}$ where $\nu_{R}^{\mu}$ denotes the dissipative coefficient. 
The four-velocity of the anomalous species coincides with the bulk velocity of the plasma and, therefore, $u_{R}^{\mu} \simeq u^{\mu}$. This 
assumption simplifies a bit the discussion and corresponds to the logic followed in this investigation (see also \cite{mg}) where 
the single fluid approach is privileged.  As in the analysis of non-anomalous plasmas (see section \ref{sec2}),
also in AMHD it is possible to discuss a multifluid approach entailing different velocities for the different species.

Let us now pause for a moment and recall the main features of the dissipative description adopted hereunder. Whenever dissipative effects are included both in the energy-momentum tensor and in the particle current the physical meaning of the four-velocity $u^{\mu}$ must be 
specified.  In the Eckart approach $u^{\mu}$ coincides with the velocity of particle transport \cite{degroot}. Conversely, in the Landau approach \cite{landau} the velocity $u^{\mu}$ coincides with the velocity of the energy transport defined by the $(0 i)$ component of the energy-momentum tensor giving the energy flux.  The Landau approach shall be privileged with the important caveat that in a perfect conductor Lorentz invariance is broken and a preferred frame (i.e. the plasma frame) arises naturally; in this frame hyperelectric fields are exactly vanishing when the conductivity goes to infinity. In the Landau approach we shall have that the global charge neutrality of the plasma is enforced by requiring that $j^{\mu} u_{\mu} =0$. If the plasma {\em is not} globally neutral, i.e. $j^{\mu} u_{\mu} = \tilde{n}$, then a second chemical potential must be introduced so that Eq. (\ref{th1a}) will be 
\begin{equation}
\varsigma \partial_{\alpha} T + \tilde{n}_{R} \partial_{\alpha} \mu_{R} + \tilde{n}\partial_{\alpha}\mu = \partial_{\alpha} p,
\label{th1b}
\end{equation} 
and $w = T \varsigma + \tilde{n} \mu + \tilde{n}_{R} \mu_{R}$. The case of a plasma which is not neutral will be treated 
in more detail in appendix \ref{APPC} with the purpose of showing that the coefficients of the magnetic and vortical currents are subjected
to a higher degree of arbitrariness as we shall discuss more precisely at the end of this section.

\subsection{Joule heating}

Equations (\ref{an1})--(\ref{th1}) can be rephrased in terms of the entropy density. 
The projection of the first expression reported in Eq. (\ref{an2}) along the four-velocity $u^{\nu}$ implies, according the results of appendix \ref{APPA}, the following relation
\begin{equation}
\nabla_{\mu} [ (p + \rho) u^{\mu}] - u^{\nu} \partial_{\nu} p - u^{\nu} \,Y_{\nu\alpha} \,j^{\alpha} =0,
\label{th2}
\end{equation}
which can be further modified by using Eq. (\ref{th1}) together with the fundamental thermodynamic identity; the result 
of this manipulation is:
\begin{equation}
\nabla_{\mu}[  \varsigma u^{\mu} - \overline{\mu}_{R} \nu_{R}^{\mu}] + \nu_{R}^{\mu}\, \partial_{\mu} \overline{\mu}_{R} +{\mathcal A}_{R}\, \overline{\mu}_{R}\, Y_{\mu\nu}\, \tilde{Y}^{\mu\nu} = 
\frac{u^{\nu}}{T} Y_{\nu\alpha}\,j^{\alpha},
\label{th3}
\end{equation}
where $\overline{\mu}_{R} = \mu_{R}/T$ denotes the rescaled chemical potential.  Equation (\ref{th3}) can be manipulated by inserting the explicit expressions of the Ohmic \cite{lic} and of the anomalous currents i.e. $j^{\alpha} = \sigma_{\mathrm{c}} Y^{\alpha\nu}\, u_{\nu} + \nu^{\alpha}$ and $j^{\alpha} = n_{R} u^{\alpha} + \nu_{R}^{\alpha}$:
\begin{equation}
\nabla_{\mu}[ \varsigma u^{\mu} - \overline{\mu}_{R} \nu_{R}^{\mu}] + \nu_{R}^{\mu} \partial_{\mu} \overline{\mu}_{R} +{\mathcal A}_{R} \overline{\mu}_{R} Y_{\mu\nu} \tilde{Y}^{\mu\nu} = \biggl(\frac{\sigma_{\mathrm{c}}}{T} \biggr) Y^{\alpha\beta} Y_{\nu\alpha} u^{\nu} u_{\beta} - \frac{\nu^{\alpha}}{T} u^{\beta} Y_{\beta\alpha}.
\label{th3a}
\end{equation}
The second law of thermodynamics implies that the covariant divergence of the entropy four-vector $\varsigma^{\mu}$
must be positive semi-definite, i.e. $\nabla_{\mu} \varsigma^{\mu} \geq 0$. Absent any anomalous current, 
the entropy of the fluid obeys $\nabla_{\mu} \varsigma^{\mu} = (\sigma_{\mathrm{c}}/T) Y_{\alpha\beta}\,Y^{\nu\alpha} u_{\nu}\,u^{\beta}$ where the term at the right hand side is the relativistic generalization of the Joule effect. This is indeed the same kind of relation 
already obtained in Eq. (\ref{prec}) of the preceding section.

The specific definition of the entropy four-vector depends on the chemical potential of the system. However, since the coefficient 
${\mathcal A}_{R}$ does not have a definite sign, the 
anomalous currents may even lead to violation of the second principle of thermodynamics (e.g. $\nabla_{\mu} \varsigma^{\mu} < 0$). Starting from a covariantly conserved total energy-momentum 
tensor without dissipative effects, the entropy four-vector is covariantly conserved . The increase of the entropy signals the presence of dissipative effects, as in the case of Joule heating. Conversely the decrease of the entropy is the result of an incomplete  definition of the entropy four-vector which is not sufficiently general, as argued in 
\cite{SS}.  Two further kinetic coefficients ${\mathcal S}_{\omega}$ and ${\mathcal S}_{B}$ will then be introduced so that 
the generalized entropy four-vector $\varsigma^{\mu}$ will become:
\begin{equation}
\varsigma^{\mu} = \varsigma u^{\mu} - \overline{\mu}_{R} \nu_{R}^{\mu} + {\mathcal S}_{\omega} \omega^{\mu} + {\mathcal S}_{B} {\mathcal B}^{\mu},
\label{ENFV}
\end{equation}
where ${\mathcal S}_{\omega}$ and ${\mathcal S}_{B}$ depend on the chemical potential and of the pressure but the arguments 
of these functions shall not be explicitly written to avoid tedious expressions. 

The vorticity four-vector $\omega^{\mu}$ appearing in Eq. (\ref{ENFV}) is defined as:
\begin{equation}
\omega^{\mu} = \tilde{f}^{\mu\alpha} u_{\alpha}, \qquad f_{\beta\gamma} = \nabla_{\beta} u_{\gamma} - \nabla_{\gamma} u_{\beta},
\label{consv}
\end{equation}
where $\tilde{f}^{\mu\alpha} = E^{\mu\alpha\beta\gamma} \, f_{\beta\gamma}/2$ is the dual tensor. In appendix \ref{APPA} and \ref{APPB} 
a collection of technical results on the general relativistic treatment of the magnetic and vortical currents has been 
included. The results reported in the appendices are by no means exhaustive and only  instrumental in easing the derivation 
of some expressions appearing hereunder. In connection with Eq. (\ref{ENFV}) it is interesting to notice that the appearance 
of the vortical current in the relativistic treatment can be physically motivated from the observation that 
the sum of the vorticity and of the magnetic field is conserved by the time evolution in flat space-time and in the 
non-relativistic limit. More specifically in an electron-ion plasma, introducing the ion mass $M$, the sum $ [(M/e) \vec{\omega} + \vec{B}]$
is conserved \cite{harr,mggrad} and this is essentially the Einstein-de Haas effect \cite{harr}. This conservation law can be generalized 
in curved space-time geometries \cite{mggrad}. Finally, inserting the entropy four-vector defined in Eq. (\ref{ENFV}) intro Eq. (\ref{th3a}) it is straightforward to obtain the following result:
\begin{eqnarray}
\nabla_{\mu} \varsigma^{\mu} - \frac{\sigma_{\mathrm{c}}}{T} Y^{\alpha\beta} Y_{\nu\alpha} u^{\nu} u_{\beta} &=& \nabla_{\mu}\biggl( {\mathcal S}_{\omega} \, \omega^{\mu}  + {\mathcal S}_{B} \,{\mathcal B}^{\mu}\biggr)
\nonumber\\
&+& \frac{\nu^{\alpha} u^{\beta}}{T} Y_{\alpha\beta} 
- \partial_{\beta} \overline{\mu}_{R} \, \nu_{R}^{\beta} - {\mathcal A}_{R}\, \overline{\mu}_{R} Y_{\alpha\beta} \tilde{Y}^{\alpha\beta}.
\label{th4}
\end{eqnarray}

\subsection{Hypermagnetic and vortical currents}
The coefficients $\nu^{\alpha}$ and $\nu^{\alpha}_{R}$ appearing in Eqs. (\ref{th3a}) and (\ref{th4}) 
must also be expressible as a combination of the vortical current and of the hypermagnetic current.  
Four different coefficients parametrize the relation between ($\nu^{\alpha}$, $\nu^{\alpha}_{R}$) and 
($\omega^{\alpha}$, ${\mathcal B}^{\alpha}$):
\begin{equation}
\nu^{\alpha} = \Lambda_{\omega} \, \omega^{\alpha} + \Lambda_{B} \, {\mathcal B}^{\alpha}, \qquad 
\nu^{\alpha}_{R} = \Lambda_{R\,\omega} \,\omega^{\alpha} + \Lambda_{R\,B} \, {\mathcal B}^{\alpha}.
\label{th5}
\end{equation}
Provided the coefficients introduced in Eq. (\ref{th5}) are specifically related to ${\mathcal S}_{\omega}$ and ${\mathcal S}_{B}$, the whole expression at the right hand side of Eq. (\ref{th4}) vanishes and the left hand side of Eq. (\ref{th4}) reproduces the standard result due to Joule heating in a conducting plasma. The relation stemming from Eq. (\ref{th4}) can be obtained with simple manipulations and it is given by:
\begin{eqnarray} 
\omega^{\alpha}\, \partial_{\alpha} {\mathcal S}_{\omega} + {\mathcal B}^{\alpha} \partial_{\alpha} {\mathcal S}_{B}
+ {\mathcal S}_{\omega} \nabla_{\alpha} \omega^{\alpha} + {\mathcal S}_{B} \nabla_{\alpha} {\mathcal B}^{\alpha} + 4 \overline{\mu}_{R} \,{\mathcal A}_{R} \, {\mathcal E}^{\alpha} {\mathcal B}_{\alpha}=
 \frac{{\mathcal E}_{\alpha} \nu^{\alpha}}{T} + \nu_{R}^{\alpha} \partial_{\alpha} \overline{\mu}_{R}.
\label{th6}
\end{eqnarray}
Exploiting the general results of Eqs. (\ref{cons6}) and (\ref{cons7}) in the case of the Ohmic current supplemented by the dissipative coefficient, the generally covariant four-divergences of $\omega^{\alpha}$ and ${\mathcal B}^{\alpha}$ are\footnote{Recall that $w$ 
denotes, in the present paper, the enthalpy density.}
\begin{eqnarray}
\nabla_{\alpha} \omega^{\alpha} &=& - \frac{2}{w} \omega^{\alpha} \partial_{\alpha} p - \frac{2}{w} \nu^{\beta} \,\omega^{\alpha} 
Y_{\alpha\beta} - \frac{2 \sigma_{\mathrm{c}}}{w} Y^{\beta\gamma} \, Y_{\alpha\beta} u_{\gamma} \omega^{\alpha} 
\label{om1}\\
\nabla_{\alpha} {\mathcal B}^{\alpha} &=& 2 Y_{\rho\sigma} \omega^{\rho} u^{\sigma} + \frac{1}{w} \tilde{Y}^{\mu\alpha} u_{\mu} \partial_{\alpha} p + \frac{\sigma_{\mathrm{c}}}{w} \tilde{Y}^{\mu\alpha}\, Y^{\beta\gamma} \,Y_{\alpha\beta} \, u_{\gamma} u_{\mu}
\nonumber\\
&+& \frac{1}{w} \tilde{Y}^{\mu\alpha} \, Y_{\alpha\beta} \, \nu^{\beta} \, u_{\mu}.
\label{om2}
\end{eqnarray}
Introducing now the fields  ${\mathcal E}^{\mu} = Y^{\mu\alpha} \, u_{\alpha}$ and ${\mathcal B}^{\mu} =  \tilde{Y}^{\mu\alpha} \, u_{\alpha}$, Eqs. (\ref{om1}) and (\ref{om2}) can be further modified:
\begin{eqnarray}
\nabla_{\alpha} \omega^{\alpha} &=& - \frac{2}{w} \omega^{\alpha} \partial_{\alpha} p - \frac{2}{w} \omega^{\alpha} {\mathcal E}_{\alpha}\,
\nu^{\beta} u_{\beta} - \frac{2}{w} u^{\rho} \, {\mathcal B}^{\sigma}\, \omega^{\alpha} [ \nu^{\beta} + \sigma_{\mathrm{c}} {\mathcal E}^{\beta}] E_{\alpha\beta\rho\sigma},
\label{om1a}\\
\nabla_{\alpha} {\mathcal B}^{\alpha} &=& 2 \omega^{\alpha} {\mathcal E}_{\alpha} - \frac{1}{w} \partial_{\alpha} p {\mathcal B}^{\alpha} 
- \frac{1}{w} u^{\beta} \nu_{\beta}\, {\mathcal E}_{\alpha} {\mathcal B}^{\alpha}.
\label{om2a}
\end{eqnarray}
Concerning Eqs. (\ref{om1a}) and (\ref{om2a}) a simple comment is in order. In the Landau approach the terms $u_{\beta} \nu^{\beta}$ and 
$u_{\alpha} \nu_{R}^{\alpha}$ vanish exactly. This is of course true also when the dissipative coefficients are defined as in Eq. (\ref{th5}) 
as it can be explicitly verified since, by definition, $u_{\beta} \omega^{\beta}$ and $u_{\beta} {\mathcal B}^{\beta}$ vanish exactly. 
Equations (\ref{om1a}) and (\ref{om2a}) can be finally inserted into Eq. (\ref{th6}) with the result that
\begin{eqnarray}
&&\omega^{\alpha} {\mathcal P}_{\alpha} + {\mathcal B}^{\alpha} {\mathcal Q}_{\alpha} + \biggl[ 2 {\mathcal S}_{B} - \biggl(\frac{\Lambda_{\omega}}{T}
\biggr)\biggr] \, \omega^{\alpha} {\mathcal B}_{\alpha}
\nonumber\\
&& + \biggl[4 \overline{\mu}_{R} {\mathcal A}_{R} - \biggl(\frac{\Lambda_{B}}{T}\biggr) \biggr]({\mathcal E}^{\alpha} {\mathcal B}_{\alpha}) 
- \frac{2}{w} \, \sigma_{\mathrm{c}}\, \omega^{\alpha} \, {\mathcal E}^{\beta}\, u^{\mu} {\mathcal B}^{\nu} \, E_{\alpha\beta\mu\nu} \, {\mathcal S}_{\omega}
=0,
\label{th7}
\end{eqnarray}
where ${\mathcal P}_{\alpha}$ and ${\mathcal Q}_{\alpha}$ are defined, respectively, as:
\begin{eqnarray}
&& {\mathcal P}_{\alpha} = \partial_{\alpha} {\mathcal S}_{\omega} - \frac{2}{w} {\mathcal S}_{\omega} \partial p - \partial_{\alpha} \overline{\mu}_{R}\,\Lambda_{R\,\omega},
\label{th8}\\
&& {\mathcal Q}_{\alpha} = \partial_{\alpha} {\mathcal S}_{B} - \frac{{\mathcal S}_{B}}{w} {\mathcal S}_{\omega} \partial_{\alpha} p - \partial_{\alpha} \overline{\mu}_{R}\, \Lambda_{R\,B}.
\label{th9}
\end{eqnarray}
The last term in Eq. (\ref{th7}) contains the explicit dependence on the conductivity. All the other terms 
of similar origin vanish because of the symmetry properties of the various currents.  
The results of Eqs. (\ref{th7})--(\ref{th9}) follow easily if we recall that, by definition, $u^{\alpha}\omega_{\alpha}$, 
$u^{\beta} {\mathcal E}_{\beta}$ and $u^{\gamma} {\mathcal B}_{\gamma}$ are all vanishing. 

In Eq. (\ref{th7}) there should be also a term containing ${\mathcal S}_{B}$ and corresponding to the one including the explicit 
dependence on $S_{\omega}$ and on the conductivity (i.e. the term proportional to $\omega^{\alpha} \, {\mathcal E}^{\beta}\, u^{\mu} {\mathcal B}^{\nu} \, E_{\alpha\beta\mu\nu}$). This term vanishes, as expected,  since it would have the same form of the last term of Eq. (\ref{th7}) but with $\omega^{\alpha}$ replaced by ${\mathcal B}^{\alpha}$: the overall coefficient will therefore 
contain the contraction of ${\mathcal B}^{\alpha} {\mathcal B}^{\nu}$ with $E_{\nu\alpha\beta\mu}$ (which is totally 
antisymmetric) so that the final contribution of this term will vanish exactly.  It is relevant to stress here that the possibility 
of a consistent analysis of the conducting case rests on the inclusion of the electric degrees of freedom. It would therefore 
be incorrect to set ${\mathcal E}^{\alpha} =0$ from the beginning since this would forbid a precise analysis of the perfectly conducting 
limit which is one of the purposes of the present investigation.

\subsection{Consistency relations}
To satisfy Eq. (\ref{th7}) the four-vectors multiplying  $\omega^{\alpha}$ and  ${\mathcal B}^{\alpha}$ 
 must vanish together with the coefficients of the terms multiplied by $\omega^{\alpha} {\mathcal B}_{\alpha}$ and ${\mathcal E}^{\alpha} {\mathcal B}_{\alpha}$. Moreover the supplementary term proportional to $\omega^{\alpha} \, {\mathcal E}^{\beta}\, u^{\mu} {\mathcal B}^{\nu} \, E_{\alpha\beta\mu\nu}$ must also vanish. To preserve the second principle of thermodynamics in a globally neutral plasma with anomalous currents and Joule heating we must 
 have that:
\begin{equation} 
{\mathcal P}_{\alpha} =0,\qquad  {\mathcal Q}_{\alpha} =0, \qquad \Lambda_{B} = 4 \mu_{R} {\mathcal A}_{R},\qquad \Lambda_{\omega} = 2 T {\mathcal S}_{B}, \qquad {\mathcal S}_{\omega} =0.
\label{th10}
\end{equation}
If, as established, ${\mathcal S}_{\omega}=0$ then Eq. (\ref{th7}) also implies that $\Lambda_{R\,\omega}=0$.
All the coefficients we ought to determine depend on $\overline{\mu}_{R}$ and on the pressure. Thus the conditions of Eq. (\ref{th10}) 
are equivalent to the following system of equations:
\begin{eqnarray}
&& \biggl( \frac{\partial {\mathcal S}_{B}}{\partial p} - \frac{S_{B}}{w} \biggr) \partial_{\alpha} p + 
\biggl( \frac{\partial {\mathcal S}_{B}}{\partial \overline{\mu}_{R}} - \Lambda_{R\,B}  \biggr) \partial_{\alpha} \overline{\mu}_{R} =0,
\label{eq1}\\
&& \Lambda_{\omega} = 2 T {\mathcal S}_{B},\qquad \Lambda_{B} = 4 {\mathcal A}_{R} \overline{\mu}_{R} T.
\label{eq2}
\end{eqnarray}
The standard thermodynamic relations giving the partial derivatives of the pressure and of the rescaled chemical 
potential with respect to the temperature are 
\begin{equation}
\biggl(\frac{\partial p}{\partial T}\biggr) = \frac{w}{T} + \tilde{n}_{R} \biggl( \frac{\partial \overline{\mu}_{R}}{\partial T}\biggr), \qquad 
\biggl(\frac{\partial \overline{\mu}_{R}}{\partial T}\biggr) = - \frac{w}{\tilde{n}_{R} T^2} + \frac{1}{\tilde{n}_{R} T} \biggl(\frac{\partial p}{\partial T}\biggr),
\label{eq3}
\end{equation}
implying that the partial derivatives of each variable with respect to the temperature (when the other variable 
is held fixed) are:
\begin{equation}
\biggl(\frac{\partial p}{\partial T}\biggr)_{\overline{\mu}_{R}} = \frac{w}{T},\qquad \biggl(\frac{\partial \overline{\mu}_{R}}{\partial T}\biggr)_{p} = - \frac{w}{\tilde{n}_{R} T^2}.
\label{eq4}
\end{equation}
With the results of Eqs. (\ref{eq3})--(\ref{eq4}), Eqs. (\ref{eq1}) and (\ref{eq2}) can be explicitly solved:
\begin{eqnarray}
&& {\mathcal S}_{B}(\overline{\mu}_{R}, T) = T \, a_{B}(\overline{\mu}_{R}),\qquad \Lambda_{R\,B} = \frac{\partial}{\partial\overline{\mu}_{R}} \biggl[ T a_{B} (\overline{\mu}_{R})\biggr],
\label{eq5}\\
&& \Lambda_{\omega}(\overline{\mu}_{R}, T) = 2 \,T^2 a_{B}(\overline{\mu}_{R}), \qquad \Lambda_{B}(\overline{\mu}_{R}, T) = 4 \,{\mathcal A}_{R} \, \overline{\mu}_{R}\, T,
\label{eq6}
\end{eqnarray}
where $a_{B}(\overline{\mu}_{R})$ is an arbitrary function of the rescaled chemical potential. Note also that 
$\Lambda_{B}$ is fully determined in terms of the coefficient of the anomaly and it is, in practice, only function 
of the chemical potential itself since, by definition, $\overline{\mu}_{R} T = \mu_{R}$. These consistency  relations will 
be discussed also in section \ref{sec5}.

All in all, the presence of an anomalous 
current induces, thanks to second principle of thermodynamics, two further terms in the Ohmic current. Starting 
with a globally neutral plasma with an anomalous current, the second principle of thermodynamics implies that the non-anomalous 
current must contain  magnetic and  vortical contributions resembling the magnetic currents induced by pseudoscalar fields. 
The induced current can be compared with the effective action for the hypercharge fields at finite fermionic density.  
In the case of right electrons ${\mathcal A}_{R} = -
g'^2 y_{R}^2 /(64 \pi^2)$ where $g'$ denotes the gauge coupling and $y_{R} = -2$ is the hypercharge assigment of the right electrons. In the comoving frame (see appendix \ref{APPB}) the interaction induced by the computed term is:
\begin{equation}
- 4 \sqrt{- g} \, \mu_{R} \, {\mathcal A}_{R}\, Y_{\mu} \tilde{Y}^{\mu\nu} \frac{g_{\nu0}}{g_{00}} = \frac{{g'}^2}{4\pi^2} \mu_{R} 
\epsilon^{i j k} Y_{i j} Y_{k}.
\label{th11}
\end{equation}
The results discussed so far refer to the globally neutral case where the current is Ohmic. If the plasma is not globally neutral the degree of arbitrariness in the determination of the consistency relations increases since the coefficients 
${\mathcal S}_{\omega}$ and ${\mathcal S}_{B}$ will  also depend on the chemical potential of the global charge 
of the plasma. This analysis is reported, for completeness, in appendix \ref{APPC} and has been also discussed, within a different 
perspective, in Ref. \cite{CC}.

\renewcommand{\theequation}{5.\arabic{equation}}
\setcounter{equation}{0}
\section{Ideal and resistive limits in AMHD}
\label{sec5}
The generally covariant discussion of the magnetic and Ohmic currents will now serve as a starting point for the analysis of conformally flat background geometries of Friedmann-Robertson-Walker type $g_{\mu\nu} = a^2(\tau) \eta_{\mu\nu}$ which are 
just slightly more restrictive than the ones discussed in section \ref{sec3}. 
The evolution equations of the system become particularly simple in terms of the rescaled electric and magnetic fields already introduced in Eqs. (\ref{div1})--(\ref{current}):
\begin{eqnarray}
&&\vec{\nabla}\cdot \vec{E} = 0, \qquad \vec{\nabla}\cdot \vec{B} =0,
\label{MX1}\\
&& \vec{\nabla} \times \vec{E}  + \partial_{\tau}\vec{B} = 0, \qquad \vec{\nabla}\times \vec{B} - \partial_{\tau}\vec{E}= 
4\pi \vec{J} + \overline{\Lambda}_{\omega} \vec{\omega} - \overline{\Lambda}_{B} \vec{B} , 
\label{MX2}
\end{eqnarray}
where the two quantities $\overline{\Lambda}_{\omega}$ and $\overline{\Lambda}_{B}$ are 
defined as $\overline{\Lambda}_{\omega}= 4 \pi a^2 \Lambda_{\omega}$ and $\overline{\Lambda}_{B} = 4 \pi a \Lambda_{B}$.
Using Eqs. (\ref{eq5}) and (\ref{eq6}) their explicit form is:
\begin{equation}
\overline{\Lambda}_{\omega} = 8\,\pi \, a^2 T^2 \, a_{B}(\overline{\mu}_{R}), \qquad \overline{\Lambda}_{B} = 16 \pi T a {\mathcal A}_{R} \overline{\mu}_{R}.
\label{MX3}
\end{equation}
From the projection of Eq. (\ref{an2}) in the direction orthogonal to $u_{\nu}$ (as discussed in Eq. (\ref{cons2}) of appendix A) the evolution equations of the bulk velocity of the plasma is given by:
\begin{eqnarray}
&& \partial_{\tau} [ W \,\vec{v}] + (\vec{v} \cdot \vec{\nabla}) [ W\, \vec{v}]  + \vec{v}\, \vec{\nabla}\cdot[ W \vec{v}] = - \vec{\nabla}P + 
\vec{J} \times \vec{B} + \eta \biggl[ \nabla^2 \vec{v} + \frac{1}{3} \vec{\nabla} (\vec{\nabla}\cdot\vec{v})\biggr],
\label{MX4}\\
&& \partial_{\tau} \epsilon + \vec{\nabla}\cdot[W \vec{v}] - \vec{E} \cdot \vec{J} =0, 
\label{MX5}
\end{eqnarray}
where $W$ denotes the rescaled enthalpy density and $(\epsilon,\,P)$ are the rescaled energy density and pressure: 
\begin{equation}
W= a^4\, w= a^4 ( p + \rho) = \epsilon + P ,\qquad P = a^4 p, \qquad \epsilon = a^4 \rho.
\label{resc}
\end{equation}
Equations (\ref{MX4}) and (\ref{MX5}) can be simplified in the case of an incompressible closure where $\vec{\nabla} \cdot \vec{v} =0$ 
even if probably this is not the most physically justified closure prior to matter radiation equality (see e. g. \cite{mg2}).
For the slow modes of the plasma the displacement current can be dropped in Eq. (\ref{MX2}) so that the generalized magnetic diffusivity equation is: 
\begin{equation}
\partial_{\tau} \vec{B}= \vec{\nabla}\times(\vec{v} \times \vec{B}) + \frac{\nabla^2 \vec{B}}{ 4 \pi \sigma} + \frac{\vec{\nabla} \times (\overline{\Lambda}_{\omega} \vec{\omega})}{4 \pi \sigma} - 
\frac{\vec{\nabla} \times(\overline{\Lambda}_{B} \vec{B})}{4 \pi \sigma}.
\label{MX6}
\end{equation}
Equation (\ref{MX6}) should be compared with Eq. (\ref{mdiff}) holding in the pseudoscalar case. 
Focussing the attention on the terms containing the conductivity we have:
\begin{equation}
\frac{ \overline{\Lambda}_{\omega}}{4 \pi \sigma} =  \frac{ \overline{T}^2 }{\sigma} a_{B}(\overline{\mu}_{R}),\qquad \frac{ \overline{\Lambda}_{B}}{4 \pi \sigma} = \frac{\overline{T} }{4 \pi \sigma} {\mathcal A}_{R} \overline{\mu}_{R}, 
\label{MX7}
\end{equation}
where $\overline{T} = a T$ denotes the comoving temperature, $\sigma = \sigma_{c} a$ is the comoving conductivity.

The rescaled chemical potential enters the infinitely conducting limit of Eq. (\ref{MX7}) since it is generally plausible that 
$\overline{\mu}_{R} \ll 1$ while $\overline{T}$ and $\sigma$ are approximately constant in time.  The smallness of the particle asymmetries 
is the rationale for the minuteness of the rescaled chemical potentials in approximate thermal equilibrium. Positing, for simplicity, 
that all the species can be treated as being ultrarelativistic at temperatures larger than a certain reference temperature 
(e.g. the temperature of the electroweak phase transition) and assuming the minimal standard 
model of electroweak interactions with three families and massless neutrinos, there are three conserved global charges supplemented 
by the hypercharge and by the third component of the weak isospin. If the plasma is hypercharge neutral  
the value of the chemical potential can be estimated from the asymmetry in the case where all the standard model 
charges are in complete thermal equilibrium \cite{mg}. If all the asymmetry is attributed to the right electrons (which is, in some 
sense, the most favourable situation) then $\overline{\mu}_{R} = (87 \pi^2/220) \, N_{\mathrm{eff}} (n_{R}/\varsigma)$ where $N_{\mathrm{eff}} = 106.75$.
This means that, indeed, $\overline{\mu}_{R} \ll 1$.

Denoting with $m$ the mass of the lightest charge carrier,  $\sigma \propto \sigma_{0} \overline{T}(1 + m\,a/\overline{T})^{-1/2}$  and $\sigma_{0}$ can be estimated on explicit models like the 
ones of Ref. \cite{cond}. In the case of an electromagnetic plasma $\sigma_{0} \propto \alpha_{\mathrm{em}}^{-1}$.  
The balance between the two terms in Eq. (\ref{MX7}) depends on the value of $a_{B}(\overline{\mu}_{R})$ (which is not fixed) but in the limit of infinite conductivity Eq. (\ref{MX6}) leads to 
\begin{equation}
\partial_{\tau} \vec{B} = \vec{\nabla}\times(\vec{v} \times \vec{B}) + {\mathcal O}\biggl(\frac{\overline{\mu}_{R}}{\sigma}\biggr),
\label{MX8}
\end{equation}
which is qualitatively similar to the result of Eq. (\ref{mdiff}).
Defining the vector potential in the Coulomb gauge, Eq. (\ref{MX8}) becomes, up to small corrections, 
$\partial_{\tau} \vec{A}= \vec{v} \times (\vec{\nabla}\times\vec{A})$. The classic analysis 
of Woltjer and Chandrasekhar \cite{woltjer} (see also \cite{fermi,kendall}) can then be exploited. The magnetic energy density shall then be minimized in a finite volume 
under the assumption of constant magnetic helicity by introducing the Lagrange multiplier $\zeta$. By taking the functional variation 
of\footnote{Following the treatment of Ref. \cite{woltjer} (see also \cite{fermi,kendall}) we assume that $V$ is the fiducial volume of a closed system. In the 
present case it could be identified, for instance, with the volume of the particle horizon at a given epoch after the end of inflation.}
\begin{equation}
{\mathcal G} =\int_{V} d^{3} x\{ |\vec{\nabla} \times \vec{A}|^2 - \zeta \vec{A} \cdot (\vec{\nabla}\times\vec{A}) \},
\label{MIN}
\end{equation}
with respect to $\vec{A}$ and by requiring $\delta {\mathcal G} =0$, the 
configurations minimizing ${\mathcal G}$ are such that $\vec{\nabla} \times \vec{B} = \zeta \vec{B}$. These configurations have been used to describe hypermagnetic knots (see \cite{mg}, third and fourth papers); in this case $\zeta$ 
with the dimensions of an inverse length and giving the scale of the hypermagnetic knot which are related to Chern-Simons waves.  Configurations with finite energy  and finite helicity can also be constructed \cite{mg,kn}.  The configurations with constant $\zeta$ represent the lowest state of magnetic energy which a closed system may attain also in the case where anomalous currents are present, provided the ambient plasma is perfectly conducting.

The limit $\sigma \to \infty$  can be corroborated by explicit solutions valid in the presence of  anomalous symmetries in conformally flat space-time geometries and minimizing, asymptotically, the functional of Eq. (\ref{MIN}). Let us now use the configurations (\ref{MIN}) and try to find solutions 
of our system. For sake of simplicity we shall assume the constancy of the rescaled enthalpy $W$ both in space and time. This means that the rescaled energy density and pressure are also constant in time provided, the plasma is dominated by radiation and $P = \epsilon/3$. For consistency the fluid should be incompressible in the absence of the relativistic fluctuations of the geometry (see however \cite{mg2}). Under these simplifying (but realistic) assumptions Eqs. (\ref{MX4}) and (\ref{MX6}) can be rewritten as:
 \begin{eqnarray}
&& \partial_{\tau} \vec{v} = \vec{v} \times \vec{\omega} - \vec{\nabla} \biggl( \frac{P}{W} + \frac{v^2}{2}\biggr) + \frac{\vec{J} \times\vec{B}}{W} + \nu_{\mathrm{kin}} \nabla^2 \vec{v}, 
\label{sim1}\\
&& \partial_{\tau} \vec{B}  = \vec{\nabla} \times (\vec{v} \times \vec{B}) + \lambda_{\omega} \vec{\nabla} \times \vec{\omega} - \lambda_{B}  \vec{\nabla} \times \vec{B} + \nu_{\mathrm{mag}} \nabla^2 \vec{B},
\label{sim2}
\end{eqnarray}
where $\nu_{\mathrm{kin}} = (\eta/W)$ and $\nu_{\mathrm{mag}} 1/(4\pi \sigma)$. Equations (\ref{sim1}) and (\ref{sim2}) 
are symmetric for a generalized self-similarity transformations 
\begin{equation}
\vec{x} \to \ell \, \vec{x}, \qquad \tau \to \ell^{1 - \delta} \tau,\qquad \vec{v} \to \ell^{\delta} \vec{v}, \qquad  \vec{B} \to \ell^{\delta} \vec{B},
\label{sim22a}
\end{equation}
holding in the so-called inertial range (i.e. when the 
magnetic forcing is absent from the right hand side of Eq. (\ref{sim1})) and 
provided ($\nu_{\mathrm{kin}}$, $\nu_{\mathrm{mag}}$)  transform as $ (\nu_{\mathrm{kin}},\,\nu_{\mathrm{mag}}) \to (\nu_{\mathrm{kin}}\,\nu_{\mathrm{mag}} )\, \ell^{1 + \delta}$. The similarity transformation of Eq. (\ref{sim22a}) holds true if  ($\lambda_{\omega}$, $\lambda_{B}$)
transform as $\lambda_{\omega} \to \lambda_{\omega} \, \ell^{\delta}$ and $\lambda_{B} \to \lambda_{B} \, \ell^{\delta}$.
Recalling that $\lambda_{\omega} \propto f_{\omega}(\overline{\mu}_{R}) \nu_{\mathrm{mag}}$ and $\lambda_{B} \propto f_{B}(\overline{\mu}_{R}) \nu_{\mathrm{mag}}$, then it also follows that 
$ f_{\omega}(\overline{\mu}_{R})$ and $ f_{B}(\overline{\mu}_{R})$ must scale as $\ell^{-1}$ if the symmetry holds true. The latter considerations 
generalize the similarity symmetry used by Olesen (see e.g. third paper of Ref. \cite{PT}) to analyze the conditions 
for inverse cascades in the standard hydromagnetic situation.

Two solutions shall now be discussed. In the first case the magnetic field is given by  $\vec{B} = \vec{H}_{0} + \vec{H}$ where $\vec{H}_{0}$ is a space-time constant while $\vec{H}$ and $\vec{v}$ are inhomogeneous and depend both on space and time. In the second case both $\vec{B}$ and $\vec{v}$ will be taken fully inhomogeneous.  Defining the auxiliary 
fields $\vec{h} = \vec{H}/\sqrt{4 \pi W}$ and $\vec{h}_{0} = \vec{H}_{0}/\sqrt{4 \pi W}$, Eqs. (\ref{sim1}) and (\ref{sim2}) are expressible as:
\begin{eqnarray}
\partial_{\tau} \vec{v} &=& \vec{v} \times (\vec{\nabla}\times \vec{v}) + (\vec{\nabla}\times \vec{h})\times\vec{h} + (\vec{h}_{0}\cdot\vec{\nabla}) \vec{h} 
\nonumber\\
&-& \sqrt{\frac{4\pi}{W}}\lambda_{\omega} \sigma (\vec{\omega} \times \vec{h}_{0} + \vec{\omega}\times \vec{h}) + 
\nu_{\mathrm{kin}} \nabla^2 \vec{v},
\label{sim1a}\\
\partial_{\tau} \vec{h} &=& \vec{\nabla}\times(\vec{v} \times \vec{h})  + (\vec{h}_{0}\cdot\vec{\nabla}) \vec{v} + \frac{\lambda_{\omega}}{\sqrt{4 \pi W}}
 \vec{\nabla}\times\vec{\omega} - \lambda_{B} \vec{\nabla}\times \vec{h} + \nu_{\mathrm{mag}} \nabla^2 \vec{h}.
\label{sim2a}
\end{eqnarray}
After careful inspection of  Eqs. (\ref{sim1a})--(\ref{sim2a}) there are two possibilities for a consistent solution. If $\lambda_{\omega}=0$ and $h_{0} \neq 0$
Eqs. (\ref{sim1a})--(\ref{sim2a}) are solved provided the functional of Eq. (\ref{MIN}) is minimized and, consequently,  $\vec{\nabla}\times\vec{h} = k \vec{h}$ and $\vec{\nabla}\times\vec{v} = k \vec{v}$. The full solution can be expressed, in a specific Cartesian coordinate system as:
\begin{equation}
\vec{v}(k,z,\tau) = v(\tau)[\cos{(k z)} \hat{e}_{x} - \sin{(k z)} \hat{e}_{y}],\qquad \vec{h}(k,z,\tau) = h(\tau)[\sin{(k z)} \hat{e}_{x} +\cos{(k z)} \hat{e}_{y}].
\label{FS}
\end{equation}
The functions $v(\tau)$ and $h(\tau)$ appearing in Eq. (\ref{FS}) must then obey:
\begin{eqnarray}
\partial_{\tau}v = k h_{0} h - \nu_{\mathrm{kin}} k^2 v, \qquad \partial_{\tau}h = - k h_{0} v - \lambda_{B} k h - \nu_{\mathrm{mag}} k^2 h.
\label{exp1}
\end{eqnarray}
As anticipated there is also a second solution of Eqs. (\ref{sim1a})--(\ref{sim2a}) which can be obtained by setting 
$h_{0} =0$ and by demanding that the velocity and the rescaled hypermagnetic field are parallel, i.e. $\vec{v}\times \vec{h} =0$ (i.e. $\vec{v} \parallel \vec{h}$). 
In the latter case, defining $\vec{v} = v(\tau) \hat{n}$ and 
 $\vec{h} = h(\tau) \hat{n}$ we have, in this second case, that 
 \begin{equation}
 \partial_{\tau} v+ k^2 \nu_{\mathrm{kin}} v=0, 
 \qquad \partial_{\tau}h + k^2 \nu_{\mathrm{mag}} h =-\frac{ k^2 \lambda_{\omega}}{\sqrt{4 \pi W}} v - k\lambda_{B} h, 
 \label{exp2}
 \end{equation}
 where, as before, $\vec{\nabla} \times \vec{h} = k \vec{h}$ and Eq. (\ref{MIN}) is minimized. Equations (\ref{exp1}) and (\ref{exp2}) can be used to investigate 
 the limit of the solutions for infinite conductivity and check that it coincides with the solution of the limit.  For instance Eq. (\ref{exp1}) in the infinite conductivity limit (i.e. $\nu_{\mathrm{mag}} \to 0$ and $\lambda_{B}\to 0$)
 for inviscid fluid (i.e. $\nu_{\mathrm{kin}} \to 0$) can be solved with the result that 
 $v(\tau) = v_{*} \cos{(k h_{0} \tau + \varphi_{*})}$ and $h(\tau) = -v_{*} \sin{(k h_{0} \tau + \varphi_{*})}$ which is 
 exactly the solution expected in the absence of anomalous currents (see e.g. last two papers in Ref. \cite{taylor}).

\newpage

\renewcommand{\theequation}{6.\arabic{equation}}
\setcounter{equation}{0}
\section{Concluding remarks}
\label{sec6}
Hydromagnetic nonlinearities in charged liquids at high magnetic Reynolds numbers lead to large-scale 
magnetic fields which are parallel rather than orthogonal to the current. Anomalous symmetries produce a similar 
effect that may even interfere with standard hydromagnetic results in a turbulent environment. Two distinct but  
equally plausible situations have been specifically scrutinized in a globally neutral system at finite conductivity: 
a plasma containing  pseudoscalar species and the anomalous currents induced by finite density effects. 

The analysis of pseudoscalar species is simplified by the covariant conservation of 
the total energy-momentum tensor of the system. The slow modes (i.e. the modes for which 
the propagation of electromagnetic disturbances is negligible) obey a generalized 
magnetic diffusivity equation where the anomalous effects are suppressed as long as 
the plasma is globally neutral, the pseudoscalar field quasi-homogeneous, and the 
conductivity parametrically large.  Instead of positing a specific action it is possible to consider the currents themselves as the building blocks of the physical 
description of the plasma. The simplest case in the framework of anomalous magnetohydrodynamics contemplates two currents one anomalous and the other non-anomalous both constrained by the canonical form of the Joule heating and by the second 
principle of thermodynamics. Supplementary terms have been shown to arise in the Ohmic current. While this treatment resembles the hydrodynamic approach to anomalous symmetries, in the present analysis, the hyperelectric 
current is not anomalous. The generalized magnetic diffusivity equation has been shown to 
include also terms proportional to the vorticity four-vector as it is intuitively plausible
by thinking of the Einstein-de Haas effect in a globally neutral plasma. The anomalous currents contribute to the evolution 
of the bulk velocity of the plasma and to generalized magnetic diffusivity equation. The perfectly conducting limit suppresses the anomalous contributions and the configurations minimizing the energy density with the constraint that the magnetic helicity be conserved coincide then with the ones obtainable in ideal magnetohydrodynamics where anomalous currents are absent. 
This observation has been used to derive hypermagnetic knot solutions in a hot plasma from their magnetic counterpart.  

\section*{Acknowledgments}
It is a pleasure to thank T. Basaglia and S. Rohr of the CERN scientific information service for their kind assistance. 

\newpage
\begin{appendix}
\renewcommand{\theequation}{A.\arabic{equation}}
\setcounter{equation}{0}
\section{Some useful generally covariant relations}
\label{APPA}
Consider a generally relativistic plasma characterized by gauge field strength $Y_{\alpha\beta}$, current $j_{\alpha}$ and four velocity $u_{\alpha}$. Using the equations of the gauge fields (i.e. $\nabla_{\mu} Y^{\mu\nu} = 4 \pi j^{\nu}$  and 
$\nabla_{\mu} \tilde{Y}^{\mu\nu} =0$) the conservation of the energy-momentum tensor 
implies that $\nabla_{\mu} T^{\mu\nu} = Y_{\nu\alpha}\, j^{\alpha}$. The latter relation can be 
projected along two orthogonal directions, i.e. $u^{\nu}$ and $h^{\alpha}_{\nu} = \delta^{\alpha}_{\nu} - u^{\alpha} u_{\nu}$ with the result:
\begin{eqnarray}
&& \nabla_{\mu} [w \, u^{\mu}] - u^{\nu} \partial_{\nu} p = Y_{\nu\alpha} \, u^{\nu}  \, j^{\alpha},
\label{cons1}\\
&& w \, u^{\mu} \nabla_{\mu} u_{\nu} - \partial_{\nu} p + u_{\nu} u^{\mu} \partial_{\mu} p = Y_{\nu\beta} j^{\beta} - Y_{\alpha\beta}\, u^{\alpha} \, j^{\beta}\, u_{\nu},
\label{cons2}
\end{eqnarray}
where $w= (\rho + p)$ denotes the enthalpy density of the fluid. The electric and magnetic fields are non-relativistic concepts while in relativistic terms the correct quantity to employ is the Maxwell 
field strength and its dual.  It is sometimes useful to decompose the gauge field strength in terms of ${\mathcal E}^{\mu}$ and ${\mathcal B}^{\mu}$:
\begin{equation}
Y_{\alpha\beta} = {\mathcal E}_{[\alpha}\, u_{\beta]}\, + \, \frac{1}{2} E_{\alpha\beta\rho\sigma}\, u^{[\rho} {\mathcal B}^{\sigma]},\qquad 
E_{\alpha\beta\rho\sigma} = \sqrt{-g}\, \epsilon_{\alpha\beta\rho\sigma},
\label{cons3}
\end{equation}
where $\epsilon_{\alpha\beta\rho\sigma}$ is the Levi-Civita symbol in 4 dimensions and 
$ {\mathcal E}_{[\alpha}\, u_{\beta]} = {\mathcal E}_{\alpha} u_{\beta} - {\mathcal E}_{\beta} u_{\alpha}$. From the definition 
of dual field strength in a four-dimensional curved space-time, i.e. $\tilde{Y}^{\mu\nu} = E^{\mu\nu\rho\sigma} Y_{\rho\sigma}/2$ we shall have, in terms of ${\mathcal E}_{\alpha}$ and ${\mathcal B}_{\beta}$:
\begin{equation}
\tilde{Y}^{\alpha\beta} = {\mathcal B}^{[\alpha} u^{\beta]}\,+\,\frac{1}{2} E^{\alpha\beta\rho\sigma}\, {\mathcal E}^{[\rho} u^{\sigma]}.
\label{cons3a}
\end{equation}
In full analogy with the gauge field strength  we can also define the vorticity four vector:
\begin{equation}
\omega^{\mu} = \tilde{f}^{\mu\alpha} u_{\alpha} \equiv \frac{1}{2} E^{\mu\alpha\beta\gamma} \, u_{\alpha}\, f_{\beta\gamma}, \qquad f_{\beta\gamma} = \nabla_{\beta} u_{\gamma} - \nabla_{\gamma} u_{\beta}.
\label{cons4}
\end{equation}
Equation (\ref{cons4})  can be inverted in terms of $f_{\gamma\beta}$ and the result is:
\begin{equation}
f_{\gamma\beta} = - E_{\gamma\beta\lambda\sigma}\, \omega^{\lambda} \,u^{\sigma} + u_{[\gamma} \, u^{\sigma} \,\nabla_{\sigma} \,u_{\beta]}.
\label{cons5}
\end{equation}
Recalling Eqs. (\ref{cons1})--(\ref{cons2}) and (\ref{cons5}) the covariant derivative of $\omega^{\mu}$ can therefore be expressed as 
\begin{equation}
\nabla_{\mu} \omega^{\mu} = - \frac{2 \omega^{\alpha}}{w} \biggl( \partial_{\alpha} p + Y_{\alpha\sigma} j^{\sigma}\biggr).
\label{cons6}
\end{equation}
In analogy with Eq. (\ref{cons6}) the covariant divergences of 
${\mathcal B}^{\mu}$ and ${\mathcal E}^{\mu}$ become:
\begin{eqnarray}
&& \nabla_{\mu} {\mathcal B}^{\mu} = 2 Y_{\rho\sigma} \, \omega^{\rho}\, u^{\sigma} + \frac{u_{\mu}\, \partial_{\alpha} p}{w} \tilde{Y}^{\mu\alpha} 
+ \frac{u_{\mu} \, Y_{\alpha\beta}}{w} \, j^{\beta} \tilde{Y}^{\mu\alpha},
\label{cons7}\\
&& \nabla_{\mu} {\mathcal E}^{\mu} = 4 \pi j^{\alpha} u_{\alpha}  - \tilde{Y}^{\mu\rho}\omega_{\mu} u_{\rho} + 
Y^{\beta\gamma} \frac{u_{\beta} \partial_{\gamma} p}{w} + \frac{Y^{\beta\gamma}u_{\beta} Y_{\gamma\alpha} j^{\alpha}}{w}.
\label{cons77}
\end{eqnarray}
In the special case where the plasma is not globally neutral and the electric current is $j^{\alpha} =  \tilde{n} u^{\alpha}$, Eqs. (\ref{cons6}) and (\ref{cons7})  become respectively 
\begin{eqnarray}
&& \nabla_{\mu} \omega^{\mu} = - 2 \frac{\omega^{\alpha} \partial_{\alpha} p}{w} - 2 \tilde{n} \frac{ {\mathcal E}^{\alpha} \omega_{\alpha}}{w},
\label{cons6a}\\
&& \nabla_{\mu} {\mathcal B}^{\mu} = 2 \, {\mathcal E}_{\alpha}\omega^{\alpha} - \frac{{\mathcal B}^{\alpha} \partial_{\alpha} p}{w} - \frac{\tilde{n}}{w}{\mathcal E}_{\alpha} {\mathcal B}^{\alpha},
\label{cons7a}\\
&& \nabla_{\mu} {\mathcal E}^{\mu} = 4 \pi \tilde{n} - \omega_{\alpha} {\mathcal B}^{\alpha} - \frac{{\mathcal E}^{\alpha}\partial_{\alpha}p}{w} -
\frac{\tilde{n} \,{\mathcal E}_{\alpha} {\mathcal E}^{\alpha}}{w}.
\label{cons77a}
\end{eqnarray}
In the absence of gauge fields, the relativistic generalization of the Helmotz equation can be written as
\begin{equation}
u^{\alpha} \nabla_{\alpha} \omega^{\mu} + \nabla_{\alpha} u^{\alpha} \omega^{\mu} - 
 \omega^{\alpha} \nabla_{\alpha} u^{\mu}  + ( u^{\alpha} \omega^{\mu} + u^{\mu} \omega^{\alpha}) 
\frac{\partial_{\alpha} p}{w} =0.
\label{cons8}
\end{equation}
\renewcommand{\theequation}{B.\arabic{equation}}
\setcounter{equation}{0}
\section{Comoving frame and physical fields}
\label{APPB}
In comoving coordinates $u_{\mu} = g_{0\mu}/\sqrt{g_{00}}$ and $u^{\mu} = \delta^{\mu}_{0}/\sqrt{g_{00}}$. In the comoving
frame the auxiliary fields defined in Eq. (\ref{cons3a}) are ${\mathcal E}^{\mu} = ( 0,\, {\mathcal E}^{i})$ and ${\mathcal B}^{\mu} = (0, {\mathcal B}^{i})$ where 
\begin{equation}
{\mathcal E}^{i} = \frac{Y^{i0}}{\sqrt{g_{00}}}, \qquad {\mathcal B}^{i} = \frac{\tilde{Y}^{i0}}{\sqrt{g_{00}}}.
\label{em1}
\end{equation}
Since ${\mathcal E}^{i}$ and ${\mathcal B}^{i}$ are not three-dimensional fields but rather the spatial 
components of a controvariant four-vector, the corresponding covariant components 
will be obviously given by ${\mathcal E}_{m} = g_{m i} \sqrt{g_{00}}\, Y^{i 0}$ 
and ${\mathcal B}_{m} = g_{m i} \sqrt{g_{00}}\, \tilde{Y}^{i 0}$.

In a perfect conductor, i.e. when the conductivity is infinite, the electric fields are completely screened. Conversely at finite 
conductivity electric fields are suppressed. In both cases Lorentz invariance is broken and it is
convenient to introduce a frame (the so called plasma frame) where the electric fields vanish. 
The spatial components of  ${\mathcal E}^{\mu}$ and ${\mathcal B}^{\mu}$ do not 
coincide with the three-dimensional fields $e^{i}$ and $b^{i}$. The three-dimensional fields can be defined 
as $Y^{i0} = g^{00} \, e^{i}$ and $Y^{ij} = - g^{00} \epsilon^{i j k}\, b_{k}$.

Since $\sqrt{- g} Y^{\mu\nu}$ and $\sqrt{- g} \tilde{Y}^{\mu\nu}$ are both invariant under Weyl rescaling, 
two Weyl invariant combinations can be introduced, i.e.  ${\overline{{\mathcal E}}}^{\mu} = \sqrt{-g} \, Y^{\mu\nu} \overline{u}_{\nu}$ 
and ${\overline{{\mathcal B}}}^{\mu} = \sqrt{-g} \, Y^{\mu\nu} \overline{u}_{\nu}$ where $\overline{u}_{\nu}$ satisfies 
$\eta_{\mu\nu} \overline{u}^{\nu}  \overline{u}^{\mu} =1$ and $\eta_{\mu\nu}$ is the Minkowski metric. The 
comoving electric and magnetic fields in three-dimensional notation are 
$\vec{E} = g^{00} \sqrt{-g} \vec{e}$ and $\vec{B} =g^{00} \sqrt{-g}\vec{b}$. Using the standard ADM decomposition 
the comoving fields are $ \vec{E} = (\sqrt{\gamma}/N) \vec{e}$ and $\vec{B} = (\sqrt{\gamma}/N) \vec{b}$ and coincide 
with the ones discussed, for instance, in \cite{mggrad}.
Following the definitions spelled out in this appendix and consistently followed in the paper we have that 
\begin{equation}
Y_{\mu\nu} \, \tilde{Y}^{\mu\nu} = 4 {\mathcal B}_{\mu} \, {\mathcal E}^{\mu} = - 4 \frac{\vec{E} \cdot\vec{B}}{\sqrt{- g}},
\label{em2}
\end{equation}
where $\sqrt{- g} = N \sqrt{\gamma} $ in the framework of the ADM decomposition. Finally, in the case 
of a conformally flat geometry we can write that the metric is  $g_{\mu\nu} = (-g)^{1/4} \eta_{\mu\nu}$ 
and the various definitions simplify so that, for instance, $\vec{E} =   (-g)^{1/4}\, \vec{e}$, $\vec{B} = 
(-g)^{1/4}\, \vec{b}$ and so on and so forth.

\renewcommand{\theequation}{C.\arabic{equation}}
\setcounter{equation}{0}
\section{The case of a non-neutral plasma}
\label{APPC}
The results of Eqs. (\ref{eq5}) and (\ref{eq6}) hold in the case of a globally neutral plasma where Ohmic and anomalous 
currents are simultaneously present. This situation will now be compared with case where, instead of a Ohmic 
current we have an ordinary particle current and the plasma is not globally neutral. In this case 
we shall have two chemical potentials one related to the anomalous current and the other related to the particle
current. The thermodynamical relations will therefore be modified and, for instance, the enthalpy density will be given by
$w = T \varsigma + \tilde{n} \mu + n_{R} \mu_{R}$. Repeating the same steps discussed before, we shall have that 
\begin{equation}
\nabla_{\alpha} [(\varsigma - \overline{\mu}_{R} - \overline{\mu}) u^{\alpha}] + \nu_{R}^{\alpha}\, \partial_{\alpha} \overline{\mu}_{R}
+ \nu^{\alpha}\, \partial_{\alpha} \overline{\mu} - 4 {\mathcal A}_{R} \overline{\mu}_{R} {\mathcal E}_{\alpha} {\mathcal B}^{\alpha} + 
\frac{\nu_{\alpha} {\mathcal E}^{\alpha}}{T} =0.
\label{nc1}
\end{equation}
The same steps outlined above can then be repeated. By defining the entropy four-vector as in Eq. (\ref{ENFV}), the 
covariant four-divergence of $\varsigma^{\mu}$ becomes:
\begin{equation}
\nabla_{\mu} \varsigma^{\mu} = \nabla_{\mu} ({\mathcal S}_{\omega} \omega^{\mu} + {\mathcal S}_{B} {\mathcal B}^{\mu} )
-  \nu_{R}^{\alpha} \partial_{\alpha} \overline{\mu}_{R} -   \nu^{\alpha} \partial_{\alpha} \overline{\mu} +
4 {\mathcal A}_{R} \overline{\mu}_{R} {\mathcal E}_{\alpha} {\mathcal B}^{\alpha} - \frac{\nu_{\alpha} {\mathcal E}^{\alpha}}{T}.
\label{nc2}
\end{equation}
We can now recall, from the general expressions of appendix \ref{APPA} and \ref{APPB} that 
\begin{eqnarray}
\nabla_{\alpha} \omega^{\alpha} &=& - \frac{2}{w} \omega^{\alpha} \partial_{\alpha} p - \frac{2}{w} \tilde{n} {\mathcal E}_{\alpha}\omega^{\alpha},
\nonumber\\
\nabla_{\alpha} {\mathcal B}^{\alpha} &=& 2 \omega^{\alpha} {\mathcal E}_{\alpha} - \frac{1}{w}\partial_{\alpha} p {\mathcal B}^{\alpha} 
- \frac{ \tilde{n}}{w} {\mathcal E}_{\alpha} {\mathcal B}^{\alpha}.
\label{nc3} 
\end{eqnarray} 
In this case the expressions of ${\mathcal P}_{\alpha}$ and ${\mathcal Q}_{\alpha}$ of Eqs. (\ref{th8}) and (\ref{th9}) become:
\begin{eqnarray}
&& {\mathcal P}_{\alpha} = \partial_{\alpha} {\mathcal S}_{\omega} - \frac{2}{w} {\mathcal S}_{\omega} \partial p - \partial_{\alpha} \overline{\mu}_{R}\,\Lambda_{R\,\omega}- \partial_{\alpha} \overline{\mu}\,\Lambda_{\omega},
\label{nc4}\\
&& {\mathcal Q}_{\alpha} = \partial_{\alpha} {\mathcal S}_{B} - \frac{{\mathcal S}_{B}}{w} {\mathcal S}_{\omega} \partial p - \partial_{\alpha} \overline{\mu}_{R}\, \Lambda_{R\,B} - \partial_{\alpha} \overline{\mu}\, \Lambda_{B}.
\label{nc5}
\end{eqnarray}
Two further conditions can be derived by requiring the the coefficients of ${\mathcal E}_{\alpha}\omega^{\alpha}$ and of 
${\mathcal E}_{\alpha}\, {\mathcal B}^{\alpha}$ vanish. The two relations are:
\begin{equation}
2 {\mathcal S}_{B} - \frac{\Lambda_{\omega}}{T} - 2 \frac{\tilde{n}}{w} {\mathcal S}_{\omega} =0, \qquad 
4 {\mathcal A}_{R} \overline{\mu}_{R} -  \frac{\Lambda_{B}}{T} - \frac{\tilde{n}}{w} {\mathcal S}_{B} =0.
\label{nc6}
\end{equation}
In this case ${\mathcal S}_{\omega}$ is not bound to vanish but, conversely, the system 
depends on a number of arbitrary functions. More precisely we have that:
\begin{eqnarray}
&& {\mathcal S}_{\omega}(T, \overline{\mu}, \overline{\mu}_{R}) = T^2 a_{\omega}(\overline{\mu},\overline{\mu}_{R}),\qquad 
 {\mathcal S}_{B}(T, \overline{\mu}, \overline{\mu}_{R}) = T a_{B}(\overline{\mu},\overline{\mu}_{R}),
\label{nc7}\\
&& \Lambda_{\omega}(T, \overline{\mu}, \overline{\mu}_{R}) = \frac{\partial}{\partial\overline{\mu}} \biggl[T^2 a_{\omega}(\overline{\mu},\overline{\mu}_{R})\biggr], \qquad  \Lambda_{B}(T, \overline{\mu}, \overline{\mu}_{R}) = \frac{\partial}{\partial\overline{\mu}} \biggl[T a_{B}(\overline{\mu},\overline{\mu}_{R})\biggr],
\label{nc8}\\
&& \Lambda_{\omega\,R}(T, \overline{\mu}, \overline{\mu}_{R}) = \frac{\partial}{\partial\overline{\mu}_{R}} \biggl[T^2 a_{\omega}(\overline{\mu},\overline{\mu}_{R})\biggr], \qquad  \Lambda_{B\,R}(T, \overline{\mu}, \overline{\mu}_{R}) = \frac{\partial}{\partial\overline{\mu}_{R}} \biggl[T a_{B}(\overline{\mu},\overline{\mu}_{R})\biggr].
\label{nc9}
\end{eqnarray}
From Eq. (\ref{nc6}) it follows that 
\begin{equation}
\frac{\partial a_{B}}{\partial \overline{\mu}} = 4 {\mathcal A}_{R} \overline{\mu}_{R}, \qquad \frac{\partial a_{\omega}}{\partial \overline{\mu}} = 2 a_{B}. 
\label{nc10}
\end{equation}
After integrating the two equations of Eq. (\ref{nc10}) we have that 
\begin{eqnarray}
&& a_{B}(\overline{\mu},\overline{\mu}_{R}) = 4 {\mathcal A}_{R} \overline{\mu}_{R} \overline{\mu} + f(\overline{\mu}_{R}),
\nonumber\\
&& a_{\omega}(\overline{\mu},\overline{\mu}_{R}) = 4 {\mathcal A}_{R} \overline{\mu}_{R} \overline{\mu}^2 + \overline{\mu} f(\overline{\mu}_{R}) + g(\overline{\mu}),
 \label{nc11}
 \end{eqnarray}
where $f(\overline{\mu}_{R})$ and $g(\overline{\mu})$ are two arbitrary functions of the corresponding arguments. In the simplest 
situation we can set both arbitrary functions to zero and, therefore, 
\begin{equation}
\Lambda_{\omega} = 8 {\mathcal A}_{R} \mu\, \mu_{R} \biggl( 1 - \frac{2 n T \overline{\mu}}{w}\biggr),\qquad 
\Lambda_{B} = 4 {\mathcal A}_{R} \mu\, \mu_{R} \biggl( 1 - \frac{ n T\overline{\mu}}{w}\biggr).
\label{nc12}
\end{equation}
In a relativistic plasma in thermal equilibrium, both corrections appearing in Eq. (\ref{nc12}) go as $\mu/T$.
\end{appendix}

\end{document}